# The C($^3$P) + NH$_3$ reaction in interstellar chemistry: II. Low temperature rate constants and modeling of NH, NH$_2$ and NH$_3$ abundances in dense interstellar clouds

Kevin M. Hickson,[1,2,]* Jean-Christophe Loison,[1,2] Jérémy Bourgalais,[3] Michael Capron,[3]

Sébastien D. Le Picard,[3] Fabien Goulay,[4] Valentine Wakelam[5,6]

[1]Université de Bordeaux, Institut des Sciences Moléculaires, UMR 5255, F-33400 Talence, France.

[2]CNRS, Institut des Sciences Moléculaires, UMR 5255, F-33400 Talence, France.

[3]Institut de Physique de Rennes, Astrophysique de Laboratoire, UMR CNRS 6251, Université de Rennes 1, Campus de Beaulieu, 35042 Rennes Cedex, France

[4]Department of Chemistry, West Virginia University, Morgantown, West Virginia 26506, USA

[5]Université de Bordeaux, Laboratoire d'Astrophysique de Bordeaux, UMR 5804, F-33270 Floirac, France.

[6]CNRS, Laboratoire d'Astrophysique de Bordeaux, UMR 5804, F-33270 Floirac, France.

*kevin.hickson@u-bordeaux.fr

**ABSTRACT**

A continuous supersonic flow reactor has been used to measure rate constants for the $C(^3P)$ + $NH_3$ reaction over the temperature range 50–296 K. $C(^3P)$ atoms were created by the pulsed laser photolysis of $CBr_4$. The kinetics of the title reaction were followed directly by vacuum ultra-violet laser induced fluorescence (VUV LIF) of $C(^3P)$ loss and through $H(^2S)$ formation. The experiments show unambiguously that the reaction is rapid at 296 K, becoming faster at lower temperatures, reaching a value of $(1.8 \pm 0.2) \cdot 10^{-10}$ cm$^3$ molecule$^{-1}$ s$^{-1}$ at 50 K. As this reaction is not currently included in astrochemical networks, its influence on interstellar nitrogen hydride abundances is tested through a dense cloud model including gas-grain interactions. In particular, the effect of the ortho-to-para ratio of $H_2$ which plays a crucial role in interstellar $NH_3$ synthesis is examined.

## 1. INTRODUCTION

The primary goal of all astrochemical models is to reproduce the abundances of observed interstellar species in a quantitative manner. In this respect, all such models are based on a network of chemical processes which describe the rates and product formation channels of all individual reactions as a function of temperature. Unfortunately, a large majority of the included reactions have not been studied either experimentally or theoretically over the appropriate temperature range. Instead, most of these parameters are simply estimated. To improve the accuracy of astrochemical models, it is crucial to measure and calculate reaction rates and product branching ratios for the most important reactions. Typically, reactions involving species which are present at high abundances are those which could have the most impact on simulations, if the rate constants are large over the appropriate temperature range.

In the gas-phase, ammonia, $NH_3$ along with the other neutral nitrogen hydride species NH and $NH_2$ originate from molecular nitrogen reactions rather than from atomic nitrogen ones due to the low reactivity of the latter towards $H_3^+$ (Bettens & Collins 1998; Herbst et al. 1987; Milligan & McEwan 2000). $N_2$ itself is produced by atomic nitrogen reactions with small neutral radicals (Daranlot et al. 2012; Le Gal et al. 2014) which experimental and theoretical studies have shown to be less efficient than previously assumed (Daranlot et al. 2012; Daranlot et al. 2013; Daranlot et al. 2011; Jorfi & Honvault 2009; Loison et al. 2014a; Ma et al. 2012). $N_2$ subsequently reacts with $He^+$ producing $N^+$ which undergoes a series of hydrogen abstraction reactions with $H_2$ beginning with

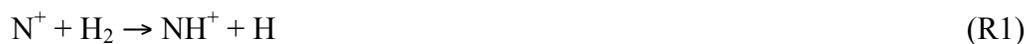

$\quad N^+ + H_2 \rightarrow NH^+ + H$ \hfill (R1)

yielding the nitrogen hydride cations $NH^+$, $NH_2^+$, $NH_3^+$ and $NH_4^+$. Dissociative recombination (DR) reactions of these species lead to the neutral hydrides NH, $NH_2$ and $NH_3$. NH is also produced from the DR reaction of $N_2H^+$ (following the $N_2 + H_3^+ \rightarrow N_2H^+ + H_2$ reaction) with a small yield (Vigren et al. 2012) which is nonetheless sufficient to make this process the major source of NH radicals in current gas-phase astrochemical models (Le Gal et al. 2014). Reaction (R1) is problematic due to the endothermicity of the $N^+$ + para-$H_2$ (p-$H_2$) reaction (Zymak et al. 2013). In the current state of modeling studies, $N^+$ reacts with ortho-$H_2$ (o-$H_2$) many orders of magnitude faster than with p-$H_2$ at temperatures relevant to dense interstellar clouds. As a result, the abundance of $NH_3$ produced by purely gas-phase chemical models is highly dependent on the $H_2$ ortho-to-para ratio (OPR) which is only poorly constrained. Using a pure gas-phase model at steady-state with an OPR for $H_2$ of $10^{-3}$, Dislaire et al. (2012) were able to reproduce the abundance ratios of NH/$NH_2$ and $NH_3$/$NH_2$ observed in IRAS 16293-2422 (Hily-Blant et al. 2010) albeit with absolute abundances an order of magnitude too small. The veracity of such model results depends entirely on the completeness of the chemical network used in simulations. The absence of key reactions and the lack of grain-surface chemistry, (a potentially important source for gas-phase NH, $NH_2$ and $NH_3$ through sequential hydrogenation reactions of N on interstellar grains followed by desorption) could drastically alter the model results.

In article I, we investigated product formation by the reaction of ground state atomic carbon $C(^3P)$ with $NH_3$. Electronic structure calculations of the intermediates, transition states and complexes along the reaction coordinate established that the reaction is characterized by a submerged barrier. Although the theoretical results show that $H_2CN$ + H, HCN + H + H and HNC + H + H are all possible exothermic product channels, experimental studies of the $H_2CN$ + H channel performed at the Advanced Light Source at 330 K coupled with measurements of the H atom product yields undertaken using the Laval nozzle technique down to 50 K clearly indicate that the H$_2$CN + H channel dominates at 300 K and below. As the electronic structure calculations suggest that the C($^3$P) + NH$_3$ reaction proceeds over a barrierless potential energy surface leading to products, this process could also be rapid at room temperature and below despite previous findings to the contrary (Deeyamulla & Husain 2007).

In the present article, we investigate the kinetics of the C($^3$P) + NH$_3$ reaction at low temperature using the supersonic flow reactor described in article I. As this process is absent from astrochemical networks, its effect on interstellar NH$_3$ abundances is currently unknown. Nevertheless, a fast C($^3$P) + NH$_3$ reaction rate at low temperature would make it one of the major interstellar ammonia loss processes. Indeed, observations of C($^3$P) by emission spectroscopy through its ground state fine structure line emission at 492 GHz (Phillips & Huggins 1981; Schilke et al. 1995) indicate that it could be one of the most abundant species in dense interstellar clouds. NH$_3$ abundances in the range 10$^{-8}$ are typically observed in dense clouds (Dickens et al. 2000; Hily-Blant et al. 2010; Ohishi et al. 1992; Ohishi & Kaifu 1998; Pratap et al. 1997). The newly measured rate constants are introduced into an astrochemical model considering nitrogen hydride formation through updated gas-phase and grain-surface chemistry. In particular, we consider the effect of the OPR ratio of H$_2$ on interstellar nitrogen hydride abundances.

## 2. EXPERIMENTAL DETAILS

A continuous supersonic flow reactor based on the original instrument of Rowe et al. (1984) was used for the present measurements. Only experimental details specific to the current investigation are outlined here. Four Laval nozzles were employed in this investigation allowing five different temperatures and densities to be obtained by using Ar or N$_2$ as carrier gases. In addition, it was possible to perform measurements at room temperature by removing the nozzle and by reducing the flow velocity. The characteristics of the supersonic flows (in particular the temperature, density and velocity) were calculated from impact pressure measurements coupled with measurements of the stagnation pressure. These values are listed in Table S1.

The multiphoton dissociation of $CBr_4$ molecules was used to create $C(^3P)$ atoms using 266 nm radiation with approximately 20 mJ of pulse energy. A column of carbon atoms of uniform density was created by aligning the photolysis laser along the flow axis. $CBr_4$ was carried into the the reactor using a small carrier gas flow over solid $CBr_4$. The gas-phase concentration of $CBr_4$ was estimated to be lower than $4 \cdot 10^{12}$ molecule $cm^{-3}$ from its saturated vapor pressure.

The rate of the $C(^3P)$ + $NH_3$ reaction was measured using two methods. Firstly, $C(^3P)$ was detected through on-resonance VUV LIF via the $2s^2 2p^2\ ^3P_2 \rightarrow 2s^2 2p3d\ ^3D_3^0$ transition at 127.755 nm using a solar blind photomultiplier tube (PMT). Test experiments performed at 50 K using the corresponding $C(^3P_0)$ and $C(^3P_1)$ transitions at 127.725 nm and 127.728 nm respectively showed that equilibrium populations in the three spin-orbit components of $C(^3P)$ were attained in less than 300 ns. A pulsed narrow band dye laser operating at 766.5 nm was frequency doubled to produce UV radiation at 383.3 nm which was focused into a cell containing 13.3 kPa of xenon and 48.0 kPa of argon for phase matching mounted on the reactor at the level of the observation axis. Third harmonic generation of the UV beam allowed us to produce tunable VUV radiation which was collimated and allowed to cross the cold supersonic flow. Secondly, some kinetic measurements were also performed by following H-atom production from the C + $NH_3$ reaction through VUV LIF at 121.567 nm as described in paper I. Previous studies (Shannon et al. 2014) have shown that excited state carbon atoms $C(^1D)$ are also produced by the multiphoton dissociation of $CBr_4$ at 266 nm, at the level of 10-15% with respect to $C(^3P)$ under similar conditions. Measurements of H atom formation kinetics were therefore only performed using nitrogen based flows, as $N_2$ is known to quench efficiently $C(^1D)$ atoms at high densities (Husain & Kirsch 1971). Kinetic measurements of $C(^3P)$ loss were performed using both argon and nitrogen carrier gases. As $C(^1D)$ atoms present in argon flows were expected to react at least as rapidly as $C(^3P)$ with $NH_3$ based on earlier studies of the relative reactivity of these two species (Shannon et al. 2014), the relaxation of $C(^1D)$ to $C(^3P)$ was expected to have a negligible influence. In addition, earlier measurements performed at 127 K on the $C(^3P)/C(^1D) + CH_3OH$ reactions in the presence of a large excess of $H_2$, (which rapidly removes $C(^1D)$ from the flow) yielded comparable second-order rate constants to the ones obtained without $H_2$, indicating the negligible influence of the presence of $C(^1D)$ atoms on $C(^3P)$ decays (Shannon et al. 2014).

VUV LIF signals from reagent $C(^3P)$ (or product $H(^2S)$) were recorded as a function of delay time between photolysis and probe lasers. For each $NH_3$ concentration, 30 datapoints were averaged for every time interval with decays consisting of at least 50 time points. To establish the baseline level, the probe laser was fired at a negative time delay with respect to the photolysis laser.

### 3. EXPERIMENTAL RESULTS

Typical kinetic decays obtained by the VUV LIF detection of $H(^2S)$ and $C(^3P)$ at 106 K and 75 K respectively are shown in Figure 1.

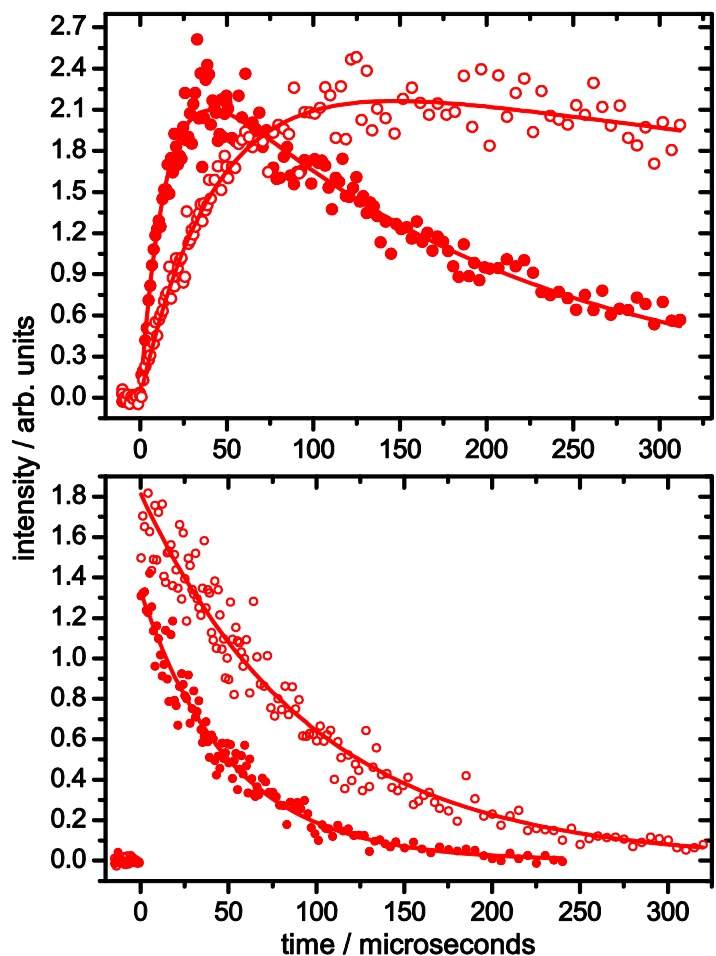

**Figure 1.** Variation of the C($^3$P) and H($^2$S) VUV LIF intensity as a function of delay time in the presence of excess NH$_3$. (Upper panel) atomic hydrogen signal at 106 K. (red filled circles) [NH$_3$] = 3.2 · 10$^{14}$ molecule cm$^{-3}$; (red open circles) [NH$_3$] = 3.2 · 10$^{13}$ molecule cm$^{-3}$. (Lower panel) atomic carbon signal at 75 K. (red filled circles) [NH$_3$] = 1.3 · 10$^{14}$ molecule cm$^{-3}$; (red open circles) [NH$_3$] = 5.3 · 10$^{13}$ molecule cm$^{-3}$. The points at negative time delays are measurements of the pre-trigger baseline values.

NH$_3$ was kept in excess of atomic carbon for all measurements allowing pseudo-first-order rate constants, $k_{1st}$, to be obtained from simple exponential fits to the carbon atom temporal profiles. $k_{1st} = k_{C+NH3}[NH_3] + k_{L(C)}$ where $k_{L(C)}$ represents any secondary losses of atomic carbon (diffusional loss and secondary reactions). For the H atom formation kinetics experiments, a biexponential function of the form

$$I_H = A\{\exp(-k_{L(H)}t) - \exp(-k_{1st}t)\} \tag{1}$$

was used to fit the H atom temporal profiles as described in paper I. None of the parameters were constrained in the biexponential fitting process.

Several NH$_3$ concentrations were used for each temperature (some measurements were also performed without NH$_3$ to assess secondary losses of C atoms) and the resulting $k_{1st}$ values were plotted against [NH$_3$]. Figure 2 shows examples of such plots with their slopes (weighted by the statistical uncertainties of individual $k_{1st}$ values) yielding rate constants for the C($^3$P) + NH$_3$ reaction at specified temperatures.

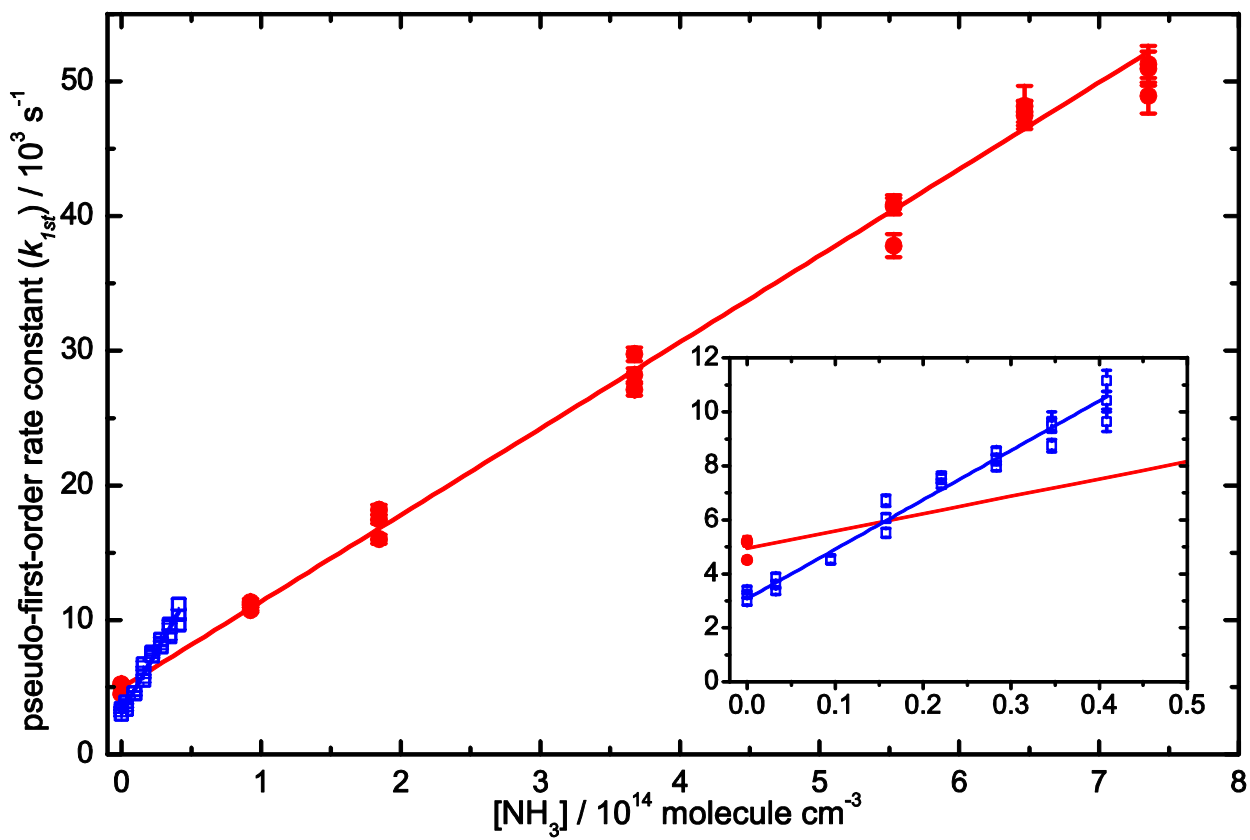

**Figure 2.** Pseudo-first-order rate constants for the C($^3$P) + NH$_3$ reaction as a function of the ammonia concentration. (**Red filled circles**) carbon atom VUV LIF measurements of the rate constant at 296 K; (blue open squares) carbon atom VUV LIF measurements of the rate constant at 50 K. (Insert) expanded view of the 50 K data. Weighted linear least squares fits to the data (**blue and red solid lines**) yielded the second-order rate constants. The error bars show the statistical uncertainty (a single standard deviation) derived from fits to kinetic decays such as those shown in Figure 1.

The exploitable ranges of NH$_3$ concentrations at 50 K and 75 K were limited by deviations in the observed $k_{1st}$ values from linearity at high NH$_3$ concentrations. As with our earlier studies of the CH + H$_2$O (Hickson et al. 2013) and C + CH$_3$OH reactions at low temperatures (Shannon et al. 2014), this effect is almost certainly due to the onset of cluster formation (NH$_3$-Ar). Only experiments conducted in the linear regime were used in the final analysis.

The temperature dependent rate constants obtained for the C($^3$P) + NH$_3$ reaction by both detection methods are presented in Figure 3 and are summarized in Table S2.

While the rate constants obtained by the H atom detection experiments are slightly larger than those obtained by the carbon atom detection method, the differences are only just outside of the combined error bars where direct comparison is possible. The results show, however, that the nature of the carrier gas does not influence the reaction rate, indicating the negligible role of complex/adduct stabilization in the C + NH$_3$ system; in good agreement with the conclusions of paper I. Nevertheless, the high values of the rate constant obtained in the present work are in stark contrast with the low value obtained by earlier experiments at 300 K (Deeyamulla & Husain 2007). These authors employed a time resolved atomic resonance absorption method to study the kinetics of the C + NH$_3$ reaction. C($^3$P) was produced by the flash photolysis of C$_3$O$_2$ molecules using a coaxial lamp at wavelengths > 160 nm and it was followed by absorption around 166 nm through several 2s$^2$2p$^2$ $^3$P → 2s$^2$2p3s $^3$P$^0$ transitions using emission from a microwave powered resonance lamp. These authors did not see any significant reaction, placing an upper limit on the rate constant value of $1.1 \cdot 10^{-11}$ cm$^3$ molecule$^{-1}$ s$^{-1}$. In contrast, we measured rate constants at 296 K of $(8.0 \pm 0.8) \cdot 10^{-11}$ cm$^3$ molecule$^{-1}$ s$^{-1}$ and $(6.4 \pm 0.6) \cdot 10^{-11}$ cm$^3$ molecule$^{-1}$ s$^{-1}$ by the H atom and C atom detection methods respectively. The noticeable differences in observed reactivity between the present work and the earlier study of Deeyamulla & Husain (2007) are difficult to reconcile. Nevertheless, we note that the range of NH$_3$ concentrations which could be practically used in this earlier study were limited due to significant absorption at 166 nm by NH$_3$ itself which may have induced large errors in the recorded intensities.

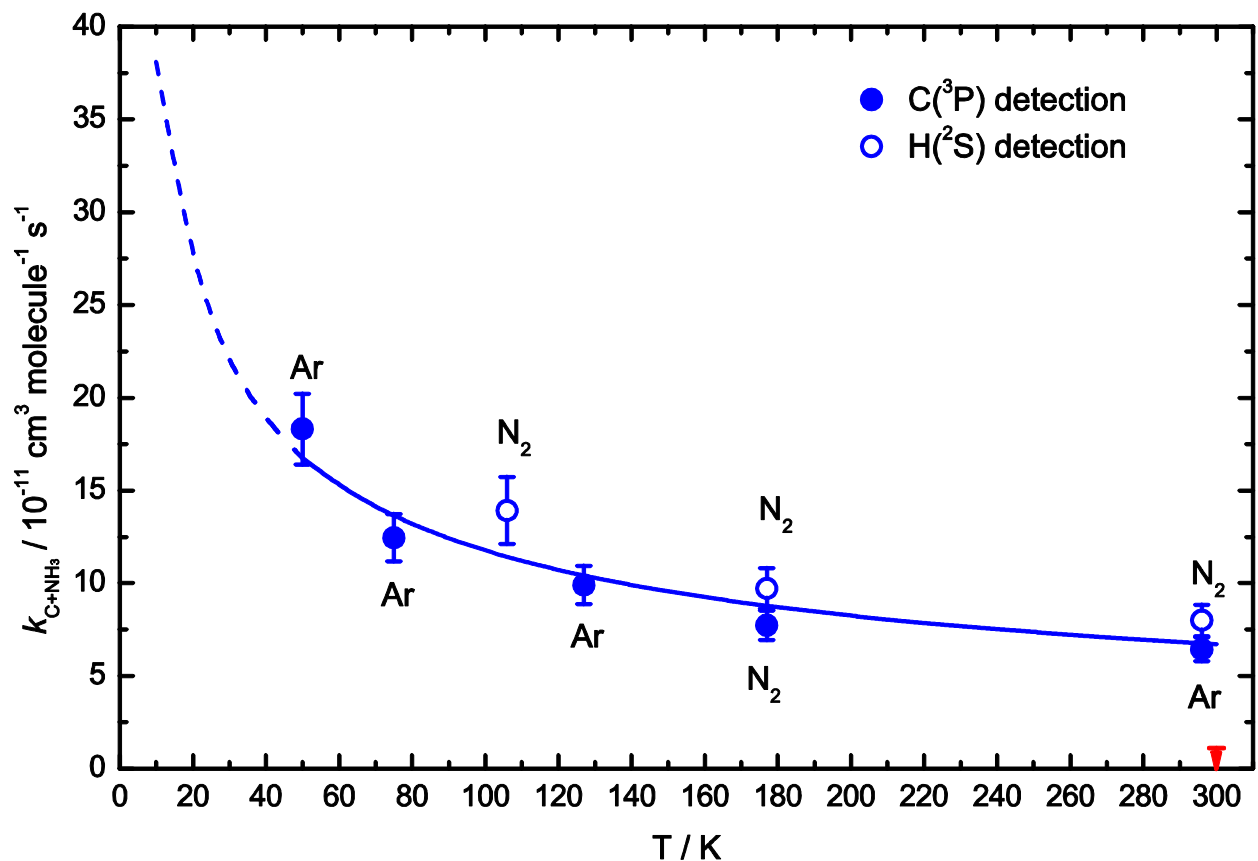

**Figure 3.** Rate constants for the C($^3$P) + NH$_3$ reaction as a function of temperature. (Blue filled circles) carbon atom VUV LIF measurements; (blue open circles) hydrogen atom VUV LIF measurements; (blue solid line) best fit to the combined carbon atom and hydrogen atom data using an equation of the form α(T/300)$^β$; (blue dashed line) extrapolation of the blue solid line to low temperature; (red line) upper limit value for the rate constant measured by Deeyamulla and Husain (2007). The labels Ar and N$_2$ reflect the specific carrier gas used for individual measurements. The error bars show the statistical uncertainty combined with a 10 % systematic uncertainty.

The reaction has a clear negative temperature dependence, with the rate constant increasing by a factor of 3 below 296 K to a value of (1.8 ± 0.2) · 10$^{-10}$ cm$^3$ molecule$^{-1}$ s$^{-1}$ at 50 K. As the experimentally determined H atom yields remain close to unity as a function of temperature and these values do not seem to vary as a function of pressure at 296 K, the low temperature reactivity seems to be governed by an increased flux through the initial transition state (TS1) leading directly to the products HCNH + H at low temperature rather than being due to stabilization of the initial van der Waals complex or adduct isomers HCNH$_2$ / H$_2$CNH (see Figure 1 of paper I for details of the potential energy surface). As a result, a simple fit to the combined atomic carbon and hydrogen data of the form $k(T) = α(T/300)^β$ can be used to represent the low pressure limiting rate constants that are appropriate for low density environments such as interstellar clouds. The fit yields the values α = (6.7 ± 2.6) · 10$^{-11}$ cm$^3$ molecule$^{-1}$ s$^{-1}$ and β = -0.51 ± 0.08 which are valid over the 50-296 K range. If these parameters are used to extrapolate the fit down to 10 K, we obtain a value for the rate constant of $k_{(C+NH3)}$(10K) = 3.8 · 10$^{-10}$ cm$^3$ molecule$^{-1}$ s$^{-1}$.

## 4. ASTROCHEMICAL MODEL

The rate constants and product branching ratios for the C + NH$_3$ reaction have been incorporated into the Nautilus model (Hersant et al. 2009; Semenov et al. 2010). The Nautilus code computes the temporal evolution of the gas-phase and ice mantle composition considering the reaction network kida.uva.2014 (http://kida.obs.u-bordeaux1.fr/models, (Wakelam et al. 2015)), the grain-surface reactions being similar to Garrod et al. (2007). The final list of reactions is composed of 9799 reactions for 710 species. We have run our model with this standard network and a modified network in which we added the C + NH$_3$ reaction and we modified the rate constant for the N$^+$ + H$_2$ reaction. Rather than considering o-H$_2$ and p-H$_2$ as two separate species (thereby requiring us to separate all reactions involving H$_2$ into their constituent o-H$_2$ and p-H$_2$ reactions), we have decided instead to focus our attention on effect of the ortho-to-para ratio (o-H$_2$/p-H$_2$) on reaction (1). To account for the difference in reactivity, we have modified the rate constant for this reaction at 10 K for two limiting values of the OPR of H$_2$. In the first case (**case I**), to simulate an OPR = 10$^{-3}$ (kinetically equilibrated H$_2$ at 10 K (Faure et al. 2013)) we use the rate constant for reaction (1) $k_{N^+ + H_2}$ (10 K) = 8.8 · 10$^{-15}$ cm$^3$ molecule$^{-1}$ s$^{-1}$ from the work of Dislaire et al. (2012) based on the experimental study of Marquette et al. (1988), reflecting the endothermic nature of the N$^+$ + p-H$_2$ reaction. In the second case (**case II**), to simulate an OPR = 3 (the statistical ratio considering the large exothermicity of H$_2$ production on grain surfaces (Gavilan et al. 2012; Naoki et al. 2010)) we use a rate constant for reaction (1) $k_{N^+ + H_2}$ (10 K) = 2.9 · 10$^{-11}$ cm$^3$ molecule$^{-1}$ s$^{-1}$, based on the recent measurements by (Zymak et al. (2013).

The initial parameters used in the simulations are listed in Table 1.

**Table 1** Elemental abundances and other model parameters

| Element | Abundance[a] | nH + 2nH$_2$ / cm$^{-3}$ | T/ K | Cosmic ray ionization | Visual |
|---|---|---|---|---|---|

|   |   |   |   | rate / s$^{-1}$ | extinction |
|---|---|---|---|---|---|
| H$_2$ | 0.5 | $2 \cdot 10^4$ | 10 | $1.3 \cdot 10^{-17}$ | 10 |
| He | 0.09 | | | | |
| C$^+$ | $1.7 \cdot 10^{-4}$ | | | | |
| N | $6.2 \cdot 10^{-5}$ | | | | |
| O | $2.4 \cdot 10^{-4}$ | | | | |
| S$^+$ | $8.0 \cdot 10^{-8}$ | | | | |
| Si$^+$ | $8.0 \cdot 10^{-9}$ | | | | |
| Fe$^+$ | $3.0 \cdot 10^{-9}$ | | | | |
| Na$^+$ | $2.0 \cdot 10^{-9}$ | | | | |
| Mg$^+$ | $7.0 \cdot 10^{-9}$ | | | | |
| P$^+$ | $2.0 \cdot 10^{-10}$ | | | | |
| Cl$^+$ | $1.0 \cdot 10^{-9}$ | | | | |
| F | $6.7 \cdot 10^{-9}$ | | | | |

[a] Relative to total hydrogen (nH + 2nH$_2$)

All elements are initially atomic in the model, except for hydrogen which is in the form of H$_2$. The C/O elemental ratio is equal to 0.7 in this study. We considered two cases: one in which we only computed the gas-phase abundances by switching all grain-surface processes off (including adsorption) and a second one in which we run the full model including gas-grain chemistry.

First we discuss the results of the pure gas-phase model. Figure 4 compares the simulated abundances using our earlier network (with a rate constant for reaction (1) at 10 K = $6.3 \cdot 10^{-12}$ cm$^3$ molecule$^{-1}$ s$^{-1}$ based on Marquette et al. (1988), being equivalent to assuming an OPR of H$_2$ = 3) with the results of the network including the C + NH$_3$ reaction. At times when atomic carbon abundances are high (less than $3 \cdot 10^5$ years), the effect of the C + NH$_3$ reaction is clearly visible resulting in a reduction in the gas-phase NH$_3$ abundance by two orders of magnitude. As a result, the observation of high NH$_3$ abundances is likely to be a strong indication that the dense cloud is relatively evolved. The importance of the newly introduced reaction depends on the amount of atomic carbon available when the abundance of NH$_3$ is high enough. Thus the conversion time of atomic carbon to CO versus the timescale for NH$_3$ formation in the gas-phase is a key parameter of these models. Assuming that half of the initial atomic carbon has already been converted to CO in the diffuse parent cloud does not change significantly the results while starting with all atomic carbon in the form of CO reduces the effect of this reaction until a few $10^5$ yr. In the gas-phase, $NH_3$ along with the other neutral nitrogen hydride species NH and $NH_2$ originate from the $N^+ + H_2$ reaction as described earlier. Consequently, when the OPR of $H_2 = 3$, nitrogen hydrides are produced efficiently. In this scenario, the pure gas-phase network can roughly reproduce the observed $NH_3$ abundances in dark clouds (Dickens et al. 2000; Hily-Blant et al. 2010; Ohishi et al. 1992; Ohishi & Kaifu 1998; Pratap et al. 1997) when the atomic carbon abundance falls below $1 \cdot 10^{-6}$ with respect to $H_2$ (at times longer than $1 \cdot 10^6$ years in Figure 4). However, this model substantially underestimates the observed NH and $NH_2$ abundances, even at steady state.

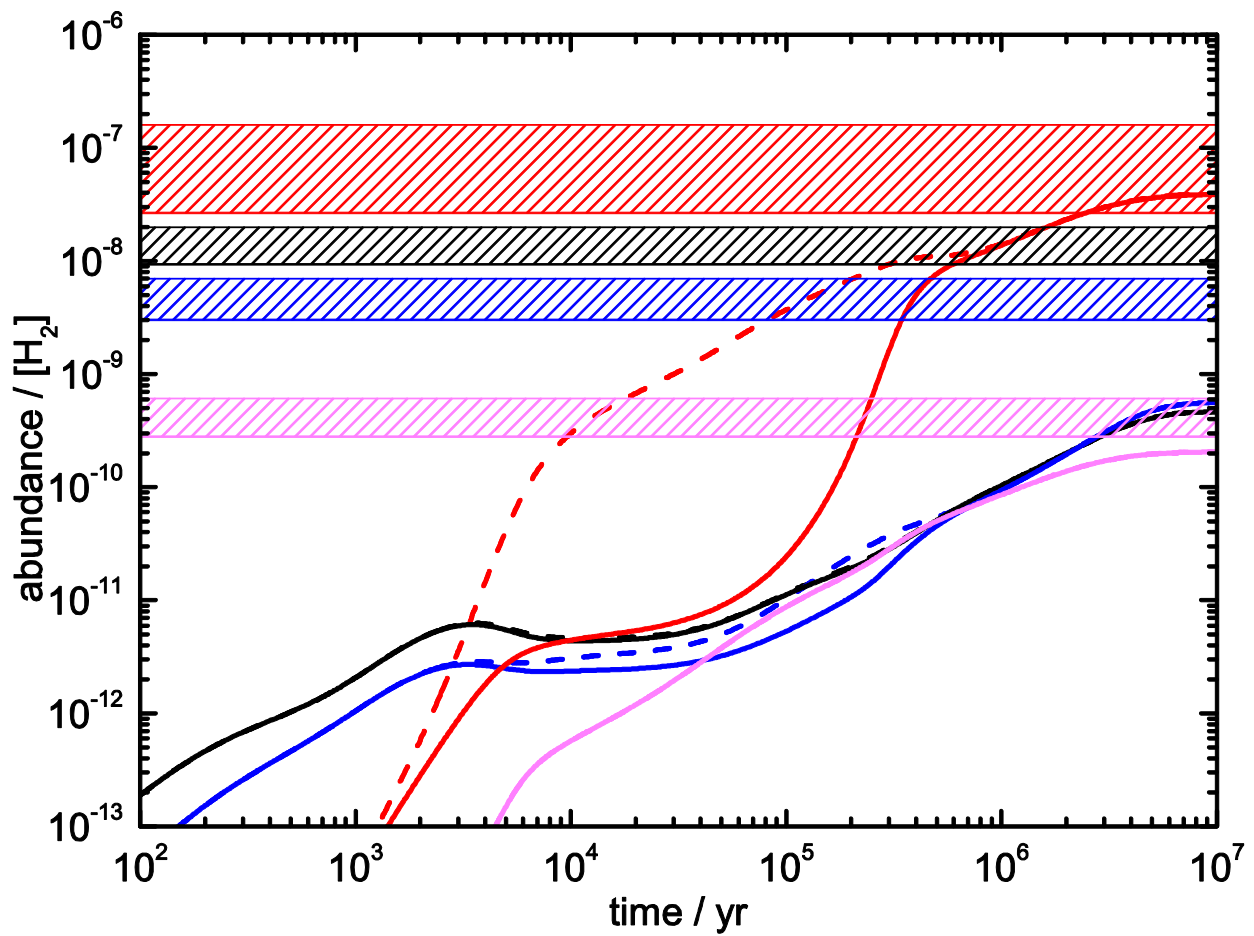

**Figure 4.** Pure gas-phase model results for the formation of selected nitrogen hydride species in dark clouds as a function of cloud age. (Dashed lines) kida.uva.2014 network. (Solid lines) the same network including the C + $NH_3$ reaction. (Hatched areas) observational constraints on the nitrogen hydride species taken from Dickens et al. (2000), Hily-Blant et al. (2010), Ohishi et al. (1992), Ohishi & Kaifu (1998), Pratap et al. (1997). $NH_3$ (Red data); $NH_2$ (blue data); NH (black data), $N_2H^+$ (pink data).

This is a notably different result compared with earlier gas-phase models (Le Gal et al. 2014), the main differences arising from the rates used for the O + NH, O + $NH_2$ and N + $NH_2$ reactions, as well as due to the newly added C + $NH_3$ reaction. The rate constants for the O + NH, O + $NH_2$ and N + $NH_2$ reactions used in this work can be found in the KIDA database and in a recent publication of datasheets from the KIDA website (Wakelam et al. 2013). The rate constant values come from various experimental studies. Adamson et al. (1994) and Hack et al. (1994) for the O + NH reaction, Adamson et al. (1994) and Dransfeld et al. (1985) for the O + $NH_2$ reaction and Dransfeld & Wagner (1987) and Whyte & Phillips (1983) for the N + $NH_2$ reaction, these rates being much higher than those used by Le Gal et al. (2014) taken from the Ohio State University (OSU) and UMIST (Woodall et al. 2007) astrochemical databases. The absence of CO depletion on grain surfaces leads to a lower $N_2H^+$ abundance in the pure gas-phase model through the $N_2H^+$ + CO reaction (slightly lower than the observed values (Dickens et al. 2000; Ohishi et al. 1992; Pratap et al. 1997)), and therefore to low NH abundances through the DR reaction of $N_2H^+$. The real OPR of $H_2$ is likely to be smaller than the statistical upper limit in dense interstellar clouds (Faure et al. 2013; Flower et al. 2006). Consequently, the use of this ratio provides only an upper limit for nitrogen hydride production. The use of an OPR value = $10^{-3}$ (**case I**) leads to even lower modeled NH and NH$_2$ abundances in addition to poor agreement between simulations and observations for NH$_3$ (see Figure 5).

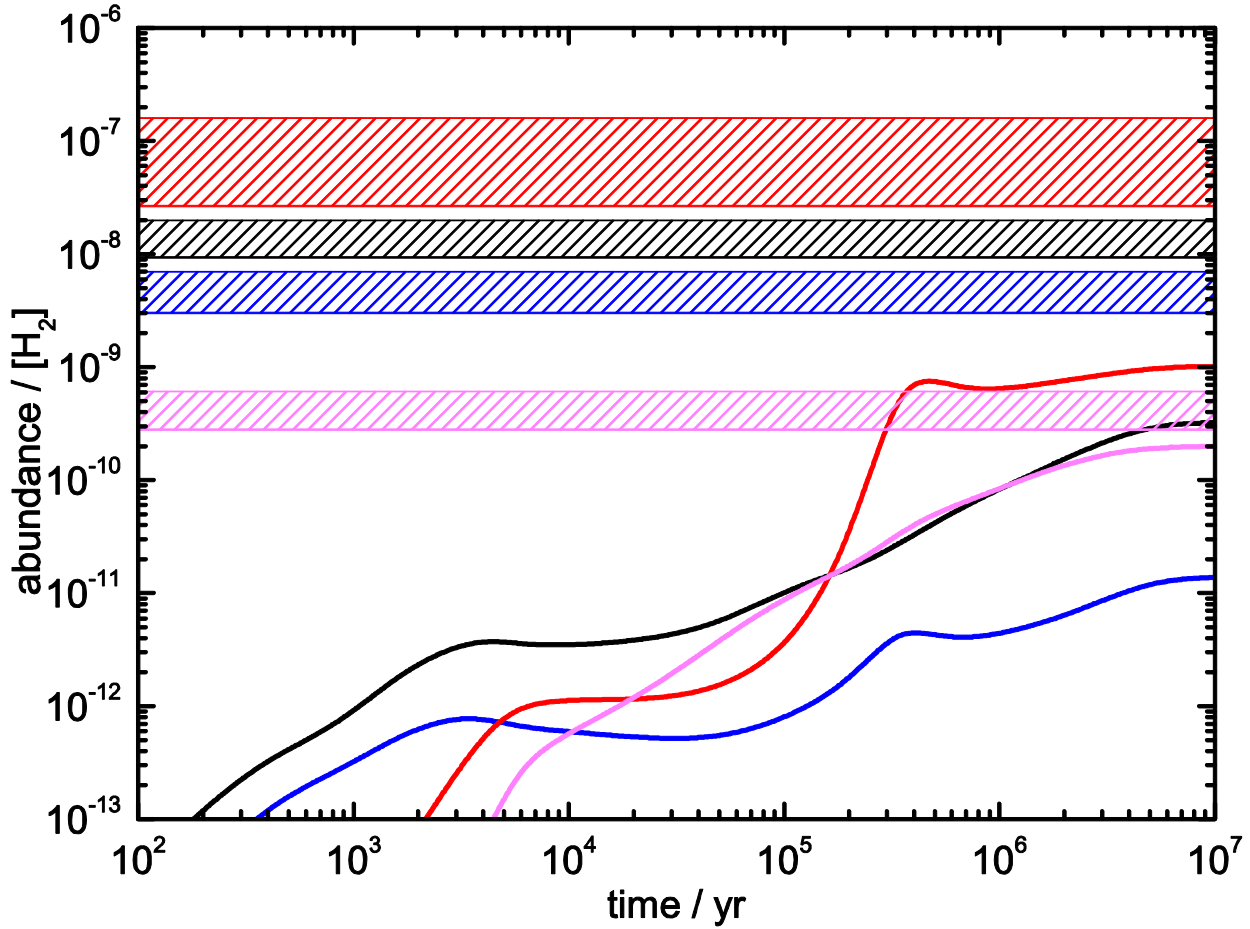

**Figure 5.** Pure gas-phase model results for the formation of selected nitrogen hydride species in dark clouds as a function of cloud age. The solid lines and hatched areas are as described in the Figure 4 caption, replacing the rate constant for reaction (1) by the **case I** limiting value.

The overall disagreement is enhanced by the inclusion of the C + NH$_3$ reaction. Indeed, for the pure gas-phase model with an OPR for H$_2$ = 10$^{-3}$, to be able to reproduce NH$_3$ abundances would require a very low atomic carbon abundance; in apparent contradiction with observations from Phillips & Huggins (1981) and Schilke et al. (1995).

We then performed simulations using the gas-grain model Nautilus where in addition to the gas-phase formation routes described above, NH, $NH_2$ and $NH_3$ are also efficiently produced through the successive hydrogenation of nitrogen atoms on grain surfaces (Hidaka et al. 2011). For each hydrogenation step, 1% of the products are considered to be released into the gas-phase due to the large reaction exothermicities (the reactive desorption mechanism). The release is efficient enough to strongly increase the gas-phase abundances of NH, $NH_2$ and $NH_3$. Despite the uncertainties on observed abundances and the fraction of molecules liberated by reactive desorption, good agreement is obtained between simulations and observations around $1 \cdot 10^6$ years as shown in Figure 6, no matter what value of the OPR of $H_2$ is used.

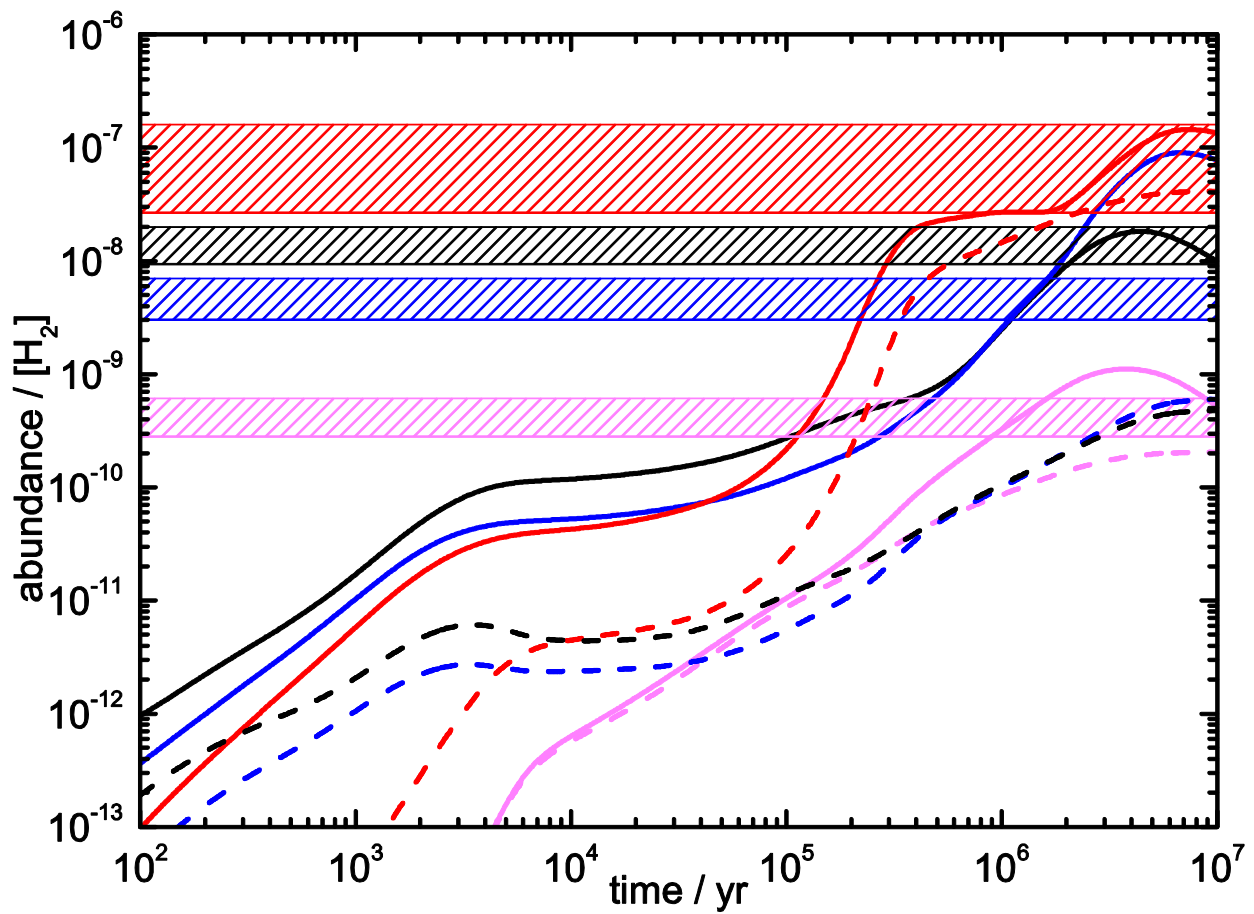

**Figure 6.** Comparison between pure gas-phase and gas-grain model results for the formation of selected nitrogen hydride species in dark clouds as a function of cloud age. Both models use the kida.uva.2014 network replacing the rate constant for reaction (1) by the **case II** limiting value. (Solid lines) gas-grain network results. (Dashed lines) pure gas-phase network results (Hatched areas) observational constraints on the nitrogen hydride species taken from Dickens et al. (2000), Hily-Blant et al. (2010), Ohishi et al. (1992), Ohishi & Kaifu (1998), Pratap et al. (1997). (Red data) $NH_3$; (blue data) $NH_2$; (black data) NH, (pink data) $N_2H^+$.

The fraction of gas-phase NH and $NH_2$ produced by grain-surface chemistry is much greater than the fraction produced by pure gas-phase chemistry itself. Indeed, with an OPR for $H_2$ = 3 (case II, which leads to the maximum gas-phase production of NH, $NH_2$ and $NH_3$), using the recent high rate constants for reaction (1) of Zymak et al. (2013), gas-phase nitrogen hydride production is still much less efficient than grain-surface production (even considering NH which is also produced in the gas-phase by the DR reaction of $N_2H^+$) as shown in Figure 6. The inclusion of grain-surface chemistry also leads to an increased gas-phase abundance of $N_2H^+$, as CO is partially depleted onto grain surfaces, thereby limiting the gas-phase $N_2H^+$ + CO reaction. This reduced $N_2H^+$ loss brings its abundance into line with observations at cloud ages where NH, $NH_2$ and $NH_3$ are also well reproduced by the gas-grain network.

The major products of the C + $NH_3$ reaction are $H_2CN$ + H as determined experimentally in article I with little or no production of HCN and/or HNC. The HCNH + H formation channel was also ruled out as HCNH was not observed in product detection experiments conducted at the Advanced Light Source synchrotron as described in article I. In order to test the effect of a small branching ratio towards HCN and HNC production on predicted HCN and HNC abundances, we set the branching ratio for the $H_2CN$ + H channel to 0.8 with the HCN + H + H and HNC + H+ H channels to 0.1 each. The simulated abundances of HCN and HNC were almost identical to the models in which the $H_2CN + H$ channel was set to 1, indicating the relatively low production of HCN and HNC compared with other sources currently contained in the model. If we examine the effect of the $C + NH_3$ reaction on $H_2CN$ production (that is comparing the gas-grain model shown in Figure 6 to the same model without the $C + NH_3$ reaction), the simulated $H_2CN$ abundance is only slightly affected. Its abundance increases by a factor of 3-4 at most over the entire cloud lifetime, the main source of $H_2CN$ being the $N + CH_3$ reaction in the current model. Indeed, simulated $H_2CN$ abundances were already significantly higher than the observed ones in dark cloud TMC-1 (Ohishi et al. 1994), possibly suggesting that other important $H_2CN$ reactions are not well described in the current network. This could be the case for the $N + CH_3$ reaction at low temperature as the rate constant for this process has a complex temperature dependence in the 200-423 K range (Marston et al. 1989).

## 5. CONCLUSION

Our new measurements of the low temperature rates of the $C + NH_3$ reaction, in addition to an updated chemistry for NH, $NH_2$ and $NH_3$ coupled with recent studies of $N_2$ production pathways (Daranlot et al. 2012; Daranlot et al. 2013; Daranlot et al. 2011; Loison et al. 2014a; Ma et al. 2012) lead to a strong disagreement between the observed and simulated abundances for NH, $NH_2$ and $NH_3$ from pure gas-phase models. In contrast, the gas-grain simulations produce good agreement with observations, a result which is essentially independent of the ortho-to-para ratio of $H_2$. Indeed, the gas-phase production of NH and $NH_2$ is predicted to be almost insignificant compared to gas-phase NH and $NH_2$ liberated from grain surfaces. As the simulated $NH_3$ abundance is low in the presence of atomic carbon (due to the rapid $C + NH_3$ reaction) we can impose a minimum dark cloud age of $3 \cdot 10^5$ years in our model (see Figure 6). Nevertheless, best agreement is obtained for a cloud age of approximately $1 \cdot 10^6$ years.

**ACKNOWLEDGEMENTS**

JCL and KMH acknowledge support from the INSU-CNRS national programs PCMI and PNP. V.W. is funded by the ERC Starting Grant (3DICE, grant agreement 336474). Acknowledgement is made to the Donors of the American Chemical Society Petroleum Research Fund for partial support of this research (PRF#53105-DN16 for FG summer support). The Rennes team acknowledges support from the Agence Nationale de la Recherche, contract ANR-11-BS04-024-CRESUSOL-01, the INSU-CNRS national program PCMI, the Institut National de Physique (INP CNRS), the Région Bretagne and the Université de Rennes 1. S.D.L.P. acknowledges financial support from the Institut Universitaire de France.

**The C($^3$P) + NH$_3$ reaction in interstellar chemistry: II. Low temperature rate constants and modeling of NH, NH$_2$ and NH$_3$ abundances in dense interstellar clouds**
*Hickson et al.*

**Online-only supplementary materials: Continuous flow characteristics and rate constants for the C($^3$P) + NH$_3$ reaction**

**Table S1** Continuous supersonic flow characteristics

| **Mach number** | **1.8 ± 0.02**[a] | **2.0 ± 0.03** | **3.0 ± 0.02** | **3.0 ± 0.1** | **3.9 ± 0.1** |
|---|---|---|---|---|---|
| Carrier gas | N$_2$ | Ar | N$_2$ | Ar | Ar |
| Density ($\cdot\ 10^{16}$ cm$^{-3}$) | 9.4 | 12.6 | 10.3 | 14.7 | 25.9 |
| Impact pressure (Pa) | 1093.2 | 1399.9 | 1786.5 | 2039.8 | 3946.3 |
| Stagnation pressure (Pa) | 1373.2 | 1853.2 | 5292.9 | 4653.0 | 15065.4 |
| Temperature (K) | 177 ± 2[a] | 127 ± 2 | 106 ± 1 | 75 ± 2 | 50 ± 1 |
| Mean flow velocity (ms$^{-1}$) | 496 ± 4[a] | 419 ± 3 | 626 ± 2 | 479 ± 3 | 505 ± 1 |

[a] The errors on the Mach number, temperature and mean flow velocity, cited at the level of one standard deviation from the mean are calculated from separate measurements of the impact pressure using a Pitot tube as a function of distance from the Laval nozzle and the stagnation pressure within the reservoir.

**Table S2** Experimental rate constants for the $C(^3P) + NH_3$ reaction obtained by following $C(^3P)$ loss and / or $H(^2S)$ formation.

| $T$ (K) | $[NH_3]^a$ | $N^b$ | $k_{C+NH3}^c$ C atom | $[NH_3]$ | $N$ | $k_{C+NH3}^c$ H atom |
|---|---|---|---|---|---|---|
| 50 ± 1 | 0 - 4.1 | 24 | (18.3 ± 1.9) | | | |
| 75 ± 2 | 1.9 - 15.3 | 35 | (12.4 ± 1.3) | | | |
| 106 ± 1 | | | | 3.2 – 38.5 | 14 | (13.9 ± 1.8) |
| 127 ± 2 | 0 – 41.0 | 37 | (9.9 ± 1.0) | | | |
| 177 ± 2 | 0 – 36.6 | 30 | (7.7 ± 0.8) | 5.3 – 65.5 | 10 | (9.7 ± 1.1) |
| 296 | 0 – 73.6 | 21 | (6.4 ± 0.6) | 8.3 – 69.4 | 18 | (8.0 ± 0.8) |

[a] / $10^{13}$ molecule cm$^{-3}$
[b] Number of individual measurements
[c] / $10^{-11}$ cm$^3$ molecule$^{-1}$ s$^{-1}$

Uncertainties on the measured rate constants represent the combined statistical (1σ) and estimated systematic (10%) errors. Uncertainties on the calculated temperatures represent the statistical (1σ) errors obtained from Pitot tube measurements of the impact pressure.